# A fast quantum mechanical algorithm for database search

Lov K. Grover
3C-404A, Bell Labs
600 Mountain Avenue
Murray Hill NJ 07974
lkgrover@bell-labs.com

## Summary

Imagine a phone directory containing $N$ names arranged in completely random order. In order to find someone's phone number with a probability of $\frac{1}{2}$, any classical algorithm (whether deterministic or probabilistic) will need to look at a minimum of $\frac{N}{2}$ names. Quantum mechanical systems can be in a superposition of states and simultaneously examine multiple names. By properly adjusting the phases of various operations, successful computations reinforce each other while others interfere randomly. As a result, the desired phone number can be obtained in only $O(\sqrt{N})$ steps. The algorithm is within a small constant factor of the fastest possible quantum mechanical algorithm.

## 1. Introduction

**1.0 Background** Quantum mechanical computers were proposed in the early 1980's [Benioff80] and in many respects, shown to be at least as powerful as classical computers - an important but not surprising result, since classical computers, at the deepest level, ultimately follow the laws of quantum mechanics. The description of quantum mechanical computers was formalized in the late 80's and early 90's [Deutsch85] [BB94] [BV93] [Yao93] and they were shown to be more powerful than classical computers on various specialized problems. In early 1994, [Shor94] demonstrated that a quantum mechanical computer could efficiently solve a well-known problem for which there was no known efficient algorithm using classical computers. This is the problem of integer factorization, i.e. finding the factors of a given integer $N$, in a time which is polynomial in $\log N$.

------------------------------------------------

This paper applies quantum computing to a mundane problem in information processing and presents an algorithm that is significantly faster than any classical algorithm can be. The problem is this: there is an unsorted database containing $N$ items out of which just one item satisfies a given condition - that one item has to be retrieved. Once an item is examined, it is possible to tell whether or not it satisfies the condition in one step. However, there does not exist any sorting on the database that would aid its selection. The most efficient classical algorithm for this is to examine the items in the database one by one. If an item satisfies the required condition stop; if it does not, keep track of this item so that it is not examined again. It is easily seen that this algorithm will need to look at an average of $\frac{N}{2}$ items before finding the desired item.

### 1.1 Search Problems in Computer Science

Even in theoretical computer science, the typical problem can be looked at as that of examining a number of different possibilities to see which, if any, of them satisfy a given condition. This is analogous to the search problem stated in the summary above, except that usually there exists some structure to the problem, i.e some sorting does exist on the database. Most interesting problems are concerned with the effect of this structure on the speed of the algorithm. For example the SAT problem asks whether it is possible to find any combination of $n$ binary variables that satisfies a certain set of clauses $C$, the crucial issue in NP-completeness is whether it is possible to solve it in time polynomial in $n$. In this case there are $N=2^n$ possible combinations which have to be searched for any that satisfy the specified property and the question is whether we can do that in a time which is polynomial in $O(\log N)$, i.e. $O(n^k)$. Thus if it were possible to reduce the number of steps to a finite power of $O(\log N)$ (instead of $O(\sqrt{N})$ as in this paper), it would yield a polynomial time algorithm for NP-complete problems.

In view of the fundamental nature of the search problem in both theoretical and applied computer sci-ence, it is natural to ask - how fast can the basic identification problem be solved without assuming anything about the structure of the problem? It is generally assumed that this limit is $O(N)$ since there are $N$ items to be examined and a classical algorithm will clearly take $O(N)$ steps. However, quantum mechanical systems can simultaneously be in multiple Schrodinger cat states and carry out multiple tasks at the same time. This paper presents an $O(\sqrt{N})$ step algorithm for the search problem.

There is a matching lower bound on how fast the desired item can be identified. [BBBV96] show in their paper that in order to identify the desired element, without any information about the structure of the database, a quantum mechanical system will need at least $\Omega(\sqrt{N})$ steps. Since the number of steps required by the algorithm of this paper is $O(\sqrt{N})$, it is within a constant factor of the fastest possible quantum mechanical algorithm.

**1.2 Quantum Mechanical Algorithms** A good starting point to think of quantum mechanical algorithms is probabilistic algorithms [BV93] (e.g. simulated annealing). In these algorithms, instead of having the system in a specified state, it is in a distribution over various states with a certain probability of being in each state. At each step, there is a certain probability of making a transition from one state to another. The evolution of the system is obtained by premultiplying this probability vector (that describes the distribution of probabilities over various states) by a state transition matrix. Knowing the initial distribution and the state transition matrix, it is possible in principle to calculate the distribution at any instant in time.

Just like classical probabilistic algorithms, quantum mechanical algorithms work with a probability distribution over various states. However, unlike classical systems, the probability vector does not completely describe the system. In order to completely describe the system we need the *amplitude* in each state which is a complex number. The evolution of the system is obtained by premultiplying this amplitude vector (that describes the distribution of amplitudes over various states) by a transition matrix, the entries of which are complex in general. The probabilities in any state are given by the square of the absolute values of the amplitude in that state. It can be shown that in order to conserve probabilities, the state transition matrix has to be unitary [BV93].

The machinery of quantum mechanical algorithms is illustrated by discussing the three operations that are needed in the algorithm of this paper. The first is the creation of a configuration in which the amplitude of the system being in any of the $2^n$ basic states of the system is equal; the second is the Walsh-Hadamard transformation operation and the third the selective rotation of different states.

A basic operation in quantum computing is that of a "fair coin flip" performed on a single bit whose states are 0 and 1 [Simon94]. This operation is represented by the following matrix: $M = \frac{1}{\sqrt{2}}\begin{bmatrix} 1 & 1 \\ 1 & -1 \end{bmatrix}$. A bit in the state 0 is transformed into a superposition in the two states: $\left(\frac{1}{\sqrt{2}}, \frac{1}{\sqrt{2}}\right)$. Similarly a bit in the state 1 is transformed into $\left(\frac{1}{\sqrt{2}}, -\frac{1}{\sqrt{2}}\right)$, i.e. the magnitude of the amplitude in each state is $\frac{1}{\sqrt{2}}$ but the *phase* of the amplitude in the state 1 is inverted. The phase does not have an analog in classical probabilistic algorithms. It comes about in quantum mechanics since the amplitudes are in general complex. In a system in which the states are described by $n$ bits (it has $2^n$ possible states) we can perform the transformation $M$ on each bit independently in sequence thus changing the state of the system. The state transition matrix representing this operation will be of dimension $2^n$ X $2^n$. In case the initial configuration was the configuration with all $n$ bits in the first state, the resultant configuration will have an identical amplitude of $2^{-\frac{n}{2}}$ in each of the $2^n$ states. This is a way of creating a distribution with the same amplitude in all $2^n$ states.

Next consider the case when the starting state is another one of the $2^n$ states, i.e. a state described by an $n$ bit binary string with some 0s and some 1s. The result of performing the transformation $M$ on each bit will be a superposition of states described by all possible $n$ bit binary strings with amplitude of each state having a magnitude equal to $2^{-\frac{n}{2}}$ and sign either + or -. To deduce the sign, observe that from the definition of the matrix $M$, i.e. $M = \frac{1}{\sqrt{2}}\begin{bmatrix} 1 & 1 \\ 1 & -1 \end{bmatrix}$, the phase of the resulting configuration is changed when a bit that was previously a 1 remains a 1 after the transformation is performed. Hence if $\bar{x}$ be the $n$-bit binary string describing the starting state and $\bar{y}$ the $n$-bit binary string describing the resulting string, the sign of the amplitude of $\bar{y}$ is determined by the parity of the bitwise dot product of $\bar{x}$ and $\bar{y}$, i.e. $(-1)^{\bar{x} \cdot \bar{y}}$. This transformation is referred to as the Walsh-Hadamard transformation [DJ92]. This operation (or a closely related operation called the Fourier Transformation) is one of the things that makes quantum mechanical algorithms more powerful than classical algorithms and forms the basis for most significant quantum mechanical algorithms.

The third transformation that we will need is the selective rotation of the phase of the amplitude in certain states. The transformation describing this for a 4 state system is of the form: 
$$\begin{bmatrix} e^{j\phi_1} & 0 & 0 & 0 \\ 0 & e^{j\phi_2} & 0 & 0 \\ 0 & 0 & e^{j\phi_3} & 0 \\ 0 & 0 & 0 & e^{j\phi_4} \end{bmatrix}, \text{ where}$$
$j = \sqrt{-1}$ and $\phi_1, \phi_2, \phi_3, \phi_4$ are arbitrary real numbers. Note that, unlike the Walsh-Hadamard transformation and other state transition matrices, the probability in each state stays the same since the square of the absolute value of the amplitude in each state stays the same.

## 2. The Abstracted Problem 

Let a system have $N = 2^n$ states which are labelled $S_1, S_2, ... S_N$. These $2^n$ states are represented as $n$ bit strings. Let there be a unique state, say $S_v$, that satisfies the condition $C(S_v) = 1$, whereas for all other states $S$, $C(S) = 0$ (assume that for any state $S$, the condition $C(S)$ can be evaluated in unit time). The problem is to identify the state $S_v$.

## 3. Algorithm

(i) Initialize the system to the distribution:
$\left( \frac{1}{\sqrt{N}}, \frac{1}{\sqrt{N}}, \frac{1}{\sqrt{N}} ... \frac{1}{\sqrt{N}} \right)$, i.e. there is the same amplitude to be in each of the $N$ states. This distribution can be obtained in $O(\log N)$ steps, as discussed in section 1.2.

(ii) Repeat the following unitary operations $O(\sqrt{N})$ times (the precise number of repetitions is important as discussed in [BBHT96]):

(a) Let the system be in any state S:
In case $C(S) = 1$, rotate the phase by $\pi$ radians;
In case $C(S) = 0$, leave the system unaltered.

(b) Apply the diffusion transform $D$ which is defined by the matrix $D$ as follows:
$D_{ij} = \frac{2}{N}$ if $i \neq j$ & $D_{ii} = -1 + \frac{2}{N}$.
This diffusion transform, $D$, can be implemented as $D = WRW$, where $R$ the rotation matrix & $W$ the Walsh-Hadamard Transform Matrix are defined as follows:
$R_{ij} = 0$ if $i \neq j$;
$R_{ii} = 1$ if $i = 0$; $R_{ii} = -1$ if $i \neq 0$.
As discussed in section 1.2:
$W_{ij} = 2^{-n/2} (-1)^{\bar{i} \cdot \bar{j}}$, where $\bar{i}$ is the binary representation of $i$, and $\bar{i} \cdot \bar{j}$ denotes the bitwise dot product of the two $n$ bit strings $\bar{i}$ and $\bar{j}$.

(iii) Sample the resulting state. In case $C(S_v) = 1$ there is a unique state $S_v$ such that the final state is $S_v$ with a probability of at least $\frac{1}{2}$.

Note that step (ii) (a) is a phase rotation transformation of the type discussed in the last paragraph of section 1.2. In a practical implementation this would involve one portion of the quantum system sensing the state and then deciding whether or not to rotate the phase. It would do it in a way so that no trace of the state of the system be left after this operation (so as to ensure that paths leading to the same final state were indistinguishable and could interfere). The implementation does *not* involve a classical measurement.

# 4. Outline of rest of paper

The loop in step (ii) above, is the heart of the algorithm. Each iteration of this loop increases the amplitude in the desired state by $O\left(\frac{1}{\sqrt{N}}\right)$, as a result in $O(\sqrt{N})$ repetitions of the loop, the amplitude and hence the probability in the desired state reach $O(1)$. In order to see that the amplitude increases by $O\left(\frac{1}{\sqrt{N}}\right)$ in each repetition, we first show that the diffusion transform, $D$, can be interpreted as an *inversion about average* operation. A simple inversion is a phase rotation operation and by the discussion in the last paragraph of section 1.2, is unitary. In the following discussion we show that the *inversion about average* operation (defined more precisely below) is also a unitary operation and is equivalent to the diffusion transform $D$ as used in step (ii)(a) of the algorithm..

Let $\alpha$ denote the average amplitude over all states, i.e. if $\alpha_i$ be the amplitude in the $i^{\text{th}}$ state, then the average is $\frac{1}{N}\sum_{i=1}^{N}\alpha_i$. As a result of the operation $D$, the amplitude in each state increases (decreases) so that after this operation it is as much below (above) $\alpha$ as it was above (below) $\alpha$ before the operation.

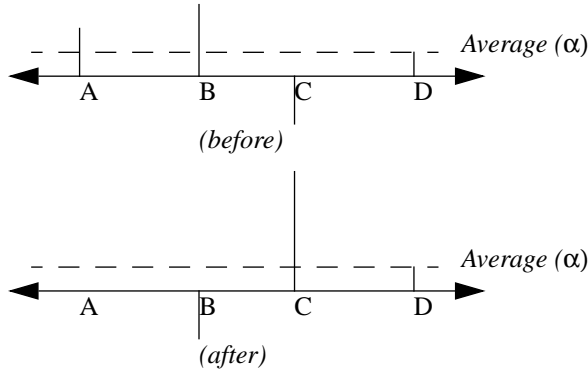

Figure 1. *Inversion about average* operation.

The diffusion transform, $D$, is defined as follows:

(4.0) $D_{ij} = \frac{2}{N}$, if $i \ne j$ & $D_{ii} = -1 + \frac{2}{N}$.

Next it is proved that $D$ is indeed the *inversion about average* as shown in figure 1 above. Observe that $D$ can be represented in the form $D \equiv -I + 2P$ where $I$ is the identity matrix and $P$ is a projection matrix with $P_{ij} = \frac{1}{N}$ for all $i, j$. The following two properties of $P$ are easily verified: first, that $P^2 = P$ & second, that $P$ acting on any vector $\bar{v}$ gives a vector each of whose components is equal to the average of all components.

Using the fact that $P^2 = P$, it follows immediately from the representation $D = -I + 2P$ that $D^2 = I$ and hence $D$ is unitary.

In order to see that $D$ is the *inversion about average*, consider what happens when $D$ acts on an arbitrary vector $\bar{v}$. Expressing $D$ as $-I + 2P$, it follows that: $D\bar{v} = (-I + 2P)\bar{v} = -\bar{v} + 2P\bar{v}$. By the discussion above, each component of the vector $P\bar{v}$ is $A$ where $A$ is the average of all components of the vector $\bar{v}$. Therefore the $i^{\text{th}}$ component of the vector $D\bar{v}$ is given by $(-v_i + 2A)$ which can be written as $(A + (A - v_i))$ which is precisely the *inversion about average*.

Next consider what happens when the *inversion about average* operation is applied to a vector where each of the components, except one, are equal to a value, say $C$, which is approximately $\frac{1}{\sqrt{N}}$; the one component that is different is negative. The average $A$ is approximately equal to $C$. Since each of the $(N-1)$ components is approximately equal to the average, it does not change significantly as a result of the inversion about average. The one component that was negative to start out, now becomes positive and its magnitude increases by approximately $2C$, which is approximately $\frac{2}{\sqrt{N}}$.

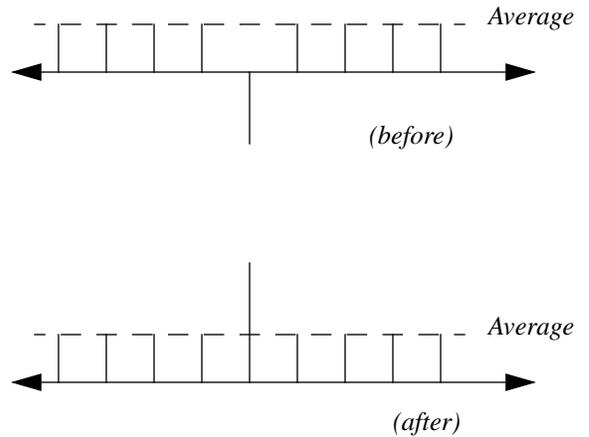

Figure 2. The *inversion about average* operation is applied to a distribution in which all but one of the com-

ponents is initially $O\left(\frac{1}{\sqrt{N}}\right)$; one of the components is initially negative.

In the loop of step (ii) of section 3, first the amplitude in a selected state is inverted (this is a phase rotation and hence a valid quantum mechanical operation as discussed in the last paragraph of section 1.2). Then the *inversion about average* operation is carried out. This increases the amplitude in the selected state in each iteration by $O\left(\frac{1}{\sqrt{N}}\right)$ (this is formally proved in the next section as theorem 3).

**Theorem 3** - Let the state vector before step (ii)(a) of the algorithm be as follows - for the one state that satisfies $C(S) = 1$, the amplitude is $k$, for each of the remaining $(N-1)$ states the amplitude is $l$ such that $\left(0 < k < \frac{1}{\sqrt{2}}\right)$ and $l > 0$. The change in $k$ ($\Delta k$) after steps (a) and (b) of the algorithm is lower bounded by $\Delta k > \frac{1}{2\sqrt{N}}$. Also after steps (a) and (b), $l > 0$.

Using theorem 3, it immediately follows that there exists a number $M$ less than $\sqrt{2N}$, such that in $M$ repetitions of the loop in step (ii), $k$ will exceed $\frac{1}{\sqrt{2}}$. Since the probability of the system being found in any particular state is proportional to the square of the amplitude, it follows that the probability of the system being in the desired state when $k$ is $\frac{1}{\sqrt{2}}$, is $k^2 = \frac{1}{2}$. Therefore if the system is now sampled, it will be in the desired state with a probability greater than $\frac{1}{2}$.

Section 6 quotes the argument from [BBBV96] that it is not possible to identify the desired record in less than $\Omega(\sqrt{N})$ steps.

## 5. Proofs

The following section proves that the system discussed in section 3 is indeed a valid quantum mechanical system and that it converges to the desired state with a probability $\Omega(1)$. It was proved in the previous section that $D$ is unitary, theorem 1 proves that it can be implemented as a sequence of three local quantum mechanical state transition matrices. Next it is proved in theorems 2 & 3 that it converges to the desired state.

As mentioned before (4.0), the diffusion transform $D$ is defined by the matrix $D$ as follows:

(5.0) $D_{ij} = \frac{2}{N}$, if $i \neq j$ & $D_{ii} = -1 + \frac{2}{N}$.

The way $D$ is presented above, it is not a local transition matrix since there are transitions from each state to all $N$ states. Using the Walsh-Hadamard transformation matrix as defined in section 3, it can be implemented as a product of three unitary transformations as $D = WRW$, each of $W$ & $R$ is a local transition matrix. $R$ as defined in theorem 2 is a phase rotation matrix and is clearly local. $W$ when implemented as in section 1.2 is a local transition matrix on each bit.

**Theorem 1** - $D$ can be expressed as $D = WRW$, where $W$, the Walsh-Hadamard Transform Matrix and $R$, the rotation matrix, are defined as follows
$R_{ij} = 0$ if $i \neq j$,
$R_{ii} = 1$ if $i = 0$, $R_{ii} = -1$ if $i \neq 0$.
$W_{ij} = 2^{-n/2}(-1)^{\bar{i}\cdot\bar{j}}$.

**Proof** - We evaluate $WRW$ and show that it is equal to $D$. As discussed in section 3, $W_{ij} = 2^{-n/2}(-1)^{\bar{i}\cdot\bar{j}}$, where $\bar{i}$ is the binary representation of $i$, and $\bar{i}\cdot\bar{j}$ denotes the bitwise dot product of the two n bit strings $\bar{i}$ and $\bar{j}$. $R$ can be written as $R = R_1 + R_2$ where $R_1 = -I$, $I$ is the identity matrix and $R_{2, 00} = 2$, $R_{2, ij} = 0$ if $i \neq 0, j \neq 0$. By observing that $MM = I$ where $M$ is the matrix defined in section 1.2, it is easily proved that $WW=I$ and hence $D_1 = WR_1W = -I$. We next evaluate $D_2 = WR_2W$. By standard matrix multiplication: $D_{2, ad} = \sum_{bc} W_{ab} R_{2, bc} W_{cd}$. Using the definition of $R_2$ and the fact $N = 2^n$, it follows that

$D_{2, ad} = 2W_{a0}W_{0d} = \frac{2}{2^n}(-1)^{\bar{a}\cdot\bar{0} + \bar{0}\cdot\bar{d}} = \frac{2}{N}$. Thus all elements of the matrix $D_2$ equal $\frac{2}{N}$, the sum of the two matrices $D_1$ and $D_2$ gives $D$.

**Theorem 2** - Let the state vector be as follows - for any one state the amplitude is $k_1$, for each of the remaining $(N-1)$ states the amplitude is $l_1$. Then after applying the diffusion transform $D$, the amplitude in the one state is $k_2 = \left(\frac{2}{N} - 1\right)k_1 + 2\frac{(N-1)}{N}l_1$ and the amplitude in each of the remaining $(N-1)$ states is $l_2 = \frac{2}{N}k_1 + \frac{(N-2)}{N}l_1$.

**Proof** - Using the definition of the diffusion transform (5.0) (at the beginning of this section), it follows that

$$k_2 = \left(\frac{2}{N} - 1\right)k_1 + 2\frac{(N-1)}{N}l_1$$

$$l_2 = \left(\frac{2}{N} - 1\right)l_1 + \frac{2}{N}k_1 + \frac{2(N-2)}{N}l_1$$

Therefore:

$$l_2 = \frac{2}{N}k_1 + \frac{(N-2)}{N}l_1$$

As is well known, in a unitary transformation the total probability is conserved - this is proved for the particular case of the diffusion transformation by using theorem 2.

**Corollary 2.1** - Let the state vector be as follows - for any one state the amplitude is $k$, for each of the remaining $(N-1)$ states the amplitude is $l$. Let $k$ and $l$ be real numbers (in general the amplitudes can be complex). Let $k$ be negative and $l$ be positive and $\left|\frac{k}{l}\right| < \sqrt{N}$. Then after applying the diffusion transform both $k_1$ and $l_1$ are positive numbers.

**Proof** - From theorem 2,

$k_1 = \left(\frac{2}{N} - 1\right)k + 2\frac{(N-1)}{N}l$. Assuming $N > 2$, it follows that $\left(\frac{2}{N} - 1\right)$ is negative; by assumption $k$ is negative and $2\frac{(N-1)}{N}l$ is positive and hence $k_1 > 0$. Similarly it follows that since by theorem 2,

$l_1 = \frac{2}{N}k + \frac{(N-2)}{N}l$, and so if the condition

$\left|\frac{k}{l}\right| < \frac{(N-2)}{2}$ is satisfied, then $l_1 > 0$. If $\left|\frac{k}{l}\right| < \sqrt{N}$, then for $N \geq 9$ the condition $\left|\frac{k}{l}\right| < \frac{(N-2)}{2}$ is satisfied and $l_1 > 0$.

**Corollary 2.2** - Let the state vector be as follows - for the state that satisfies $C(S) = 1$, the amplitude is $k$, for each of the remaining $(N-1)$ states the amplitude is $l$. Then if after applying the diffusion transformation $D$, the new amplitudes are respectively $k_1$ and $l_1$ as derived in theorem 2, then

$k_1^2 + (N-1)l_1^2 = k^2 + (N-1)l^2$.

**Proof** - Using theorem 2 it follows that

$$k_1^2 = \frac{(N-2)^2}{N^2}k^2 + 4\frac{(N-1)^2}{N^2}l^2$$
$$-\frac{4(N-2)(N-1)}{N^2}kl$$

Similarly

$$(N-1)l_1^2 = \frac{4(N-1)^2}{N^2}k^2$$
$$+ \frac{(N-2)^2}{N^2}(N-1)l^2 + \frac{4(N-2)(N-1)}{N^2}kl \ .$$

Adding the previous two equations the corollary follows.

**Theorem 3** - Let the state vector before step (a) of the algorithm be as follows - for the one state that satisfies $C(S) = 1$, the amplitude is $k$, for each of the remaining $(N-1)$ states the amplitude is $l$ such that $\left(0 < k < \frac{1}{\sqrt{2}}\right)$ and $l > 0$. The change in $k$ $(\Delta k)$ after steps (a) and (b) of the algorithm is lower bounded by $\Delta k > \frac{1}{2\sqrt{N}}$. Also after steps (a) and (b), $l > 0$.

**Proof** - Denote the initial amplitudes by $k$ and $l$, the amplitudes after the phase inversion (step (a)) by $k_1$ and $l_1$ and after the diffusion transform (step (b)) by $k_2$ and $l_2$. Using theorem 2, it follows that:

$k_2 = \left(1 - \frac{2}{N}\right)k + 2\left(1 - \frac{1}{N}\right)l$. Therefore

(5.1) $\quad \Delta k = k_2 - k = -\frac{2k}{N} + 2\left(1 - \frac{1}{N}\right)l$.

Since $\left(0 < k < \frac{1}{\sqrt{2}}\right)$, it follows from corollary 2.2 that

$|l| > \frac{1}{\sqrt{2N}}$ and since by the assumption in this theorem, $l$ is positive, it follows that $l > \frac{1}{\sqrt{2N}}$. Therefore by (5.1),

assuming non-trivial $N$, it follows that $\Delta k > \frac{1}{2\sqrt{N}}$.

In order to prove $l_2 > 0$, observe that after the phase inversion (step (a)), $k_1 < 0$ & $l_1 > 0$. Furthermore it follows from the facts $\left(0 < k < \frac{1}{\sqrt{2}}\right)$ & $|l| > \frac{1}{\sqrt{2N}}$ (discussed in the previous paragraph) that $\left|\frac{k_1}{l_1}\right| < \sqrt{N}$. Therefore by corollary 2.1, $l_2$ is positive.

# 6. How fast is it possible to find the desired element?

There is a matching lower bound from the paper [BBBV96] that suggests that it is not possible to identify the desired element in fewer than $\Omega(\sqrt{N})$ steps. This result states that any quantum mechanical algorithm running for $T$ steps is only sensitive to $O(T^2)$ queries (i.e. if there are more possible queries, then the answer to at least one can be flipped without affecting the behavior of the algorithm). So in order to correctly decide the answer which is sensitive to $N$ queries will take a running time of $T = \Omega(\sqrt{N})$. To see this assume that $C(S) = 0$ for all states and the algorithm returns the right result, i.e. that no state satisfies the desired condition. Then, by [BBBV96] if $T < \Omega(\sqrt{N})$, the answer to at least one of the queries about $C(S)$ for some $S$ can be flipped without affecting the result, thus giving an incorrect result for the case in which the answer to the query was flipped.

[BBHT96] gives a direct proof of this result along with tight bounds showing the algorithm of this paper is within a few percent of the fastest possible quantum mechanical algorithm.

# 7. Implementation considerations

This algorithm is likely to be simpler to implement as compared to other quantum mechanical algorithms for the following reasons:

(i) The only operations required are, first, the Walsh-Hadamard transform, and second, the conditional phase shift operation both of which are relatively easy as compared to operations required for other quantum mechanical algorithms [BCDP96].

(ii) Quantum mechanical algorithms based on the Walsh-Hadamard transform are likely to be much simpler to implement than those based on the "large scale Fourier transform".

(iii) The conditional phase shift would be much easier to implement if the algorithm was used in the mode where the function at each point was computed rather than retrieved form memory. This would eliminate the storage requirements in quantum memory.

(iv) In case the elements had to be retrieved from a table (instead of being computed as discussed in (iii)), in principle it should be possible to store the data in classical memory and only the sampling system need be quantum mechanical. This is because only the system under consideration needs to undergo quantum mechanical interference, not the bits in the memory. What is needed, is a mechanism for the system to be able to *feel* the values at the various datapoints something like what happens in *interaction-free measurements* as discussed in more detail in the first paragraph of the following section. Note that, in any variation, the algorithm must be arranged so as not to leave any trace of the path followed in the classical system or else the system would not undergo quantum mechanical interference.

# 8. Other observations

1. It is possible for quantum mechanical systems to make *interaction-free measurements* by using the duality properties of photons [EV93] [KWZ96]. In these the presence (or absence) of an object can be deduced by allowing for a very small probability of a photon interacting with the object. Therefore most probably the photon will not interact, however, just allowing a small probability of interaction is enough to make the measurement. This suggests that in the search problem also, it might be possible to find the object without examining all the objects but just by allowing a certain probability of examining the desired object which is something like what happens in the algorithm in this paper.

2. As mentioned in the introduction, the search algorithm of this paper does not use any knowledge about the problem. There exist fast quantum mechanical algorithms that make use of the structure of the problem at hand, e.g. Shor's factorization algorithm [Shor94]. It might be possible to combine the search scheme of this paper with [Shor94] and other quantum mechanical algorithms to design faster algorithms. Alternatively, it might be possible to combine it with efficient database search algorithms that make use of specific properties of the database. [DH96] is an example of such a recent application. [Median96] applies phase shifting techniques, similar to this paper, to develop a fast algorithm for the median estimation problem.

3. The algorithm as discussed here assumes a unique state that satisfies the desired condition. It can be easily modified to take care of the case when there are multiple states satisfying the condition $C(S) = 1$ and it is required to find one of these. Two ways of achieving this are:

(i) The first possibility would be to repeat the experiment so that it checks for a range of degeneracy, i.e. redesign the experiment so that it checks for the degeneracy of the solution being in the range $(k, k+1, \ldots 2k)$ for various $k$. Then within $\log N$ repetitions of this procedure, one can ascertain whether or not there exists at least one out of the $N$ states that satisfies the condition. [BBHT96] discusses this in detail.

(ii) The other possibility is to slightly perturb the problem in a random fashion as discussed in [MVV87] so that with a high probability the degeneracy is removed. There is also a scheme discussed in [VV86] by which it is possible to modify any algorithm that solves an NP-search problem with a unique solution and use it to solve an NP-search problem in general.

## 9. Acknowledgments